# Comparison of the *h*-index for Different Fields of Research Using Bootstrap Methodology

C.C. Malesios[1] and S. Psarakis[2]

[1] *(Corresponding Author) Department of Agricultural Development, Democritus University of Thrace, Pantazidou 193, Orestiada, e-mail: malesios@agro.duth.gr*

[2] *Department of Statistics, Athens University of Economics and Business, 76 Patission St, 10434 Athens, Greece*

## Abstract

An important disadvantage of the *h*-index is that typically it cannot take into account the specific field of research of a researcher. Usually sample point estimates of the average and median *h*-index values for the various fields are reported that are highly variable and dependent of the specific samples and it would be useful to provide confidence intervals of prediction accuracy. In this paper we apply the non-parametric bootstrap technique for constructing confidence intervals for the *h*-index for different fields of research. In this way no specific assumptions about the distribution of the empirical *h*-index are required as well as no large samples since that the methodology is based on resampling from the initial sample. The results of the analysis showed important differences between the various fields. The performance of the bootstrap intervals for the mean and median *h*-index for most fields seems to be rather satisfactory as revealed by the performed simulation.

## 1. Introduction

*Hirsch (2005)* introduced an indicator for the assessment of the research performance of scientists. The *h*-index is intended to measure simultaneously the quality and sustainability of scientific output, as well as, to some extent, the diversity of scientific research. The specific index attracted interest immediately and received great attention both in the physics community and the scientometrics literature. Not only the *h*-index has found a wide use in a very short time, but also a series of articles were subsequently published either proposing modifications of the original *h*-index for its improvement, or implementations of the proposed index.

Almost immediately, attempts to investigate the theoretical properies of the *h*-index have appeared in the literature. However, in comparison to the bulk of work on the empirical *h*-index and its applications, relatively little work has been done on the study of the theoretical *h*-index as a statistical function, allowing to construct confidence intervals, test hypotheses of interest and check validity of its statistical properties. For instance, *Glänzel (2006)* attempted to interpret theoretically some properties of the *h*-index, having assumed a citation distribution, using the theory of extreme values. Specifically, he analyzed the basic properties of the *h*-index on the basis of a probability distribution model (specifically using the Pareto distribution). *Glänzel (2006)* defined the theoretical *h*-index (which he denotes by *H*), using Gumbel's characteristic extreme values (*Gumbel 1958*). *Schubert and Glänzel (2007)* test the theoretical model of *Glänzel (2006)* in practical implementations using journal citation data, collected from the web of science database. They concluded that the theoretical Paretian model fitted perfectly to the data collected from journals. *Burrell (2007a)* also proposed a simple stochastic model in order to investigate the *h*-index and its properties.

Recently, *Beirlant and Einmahl (2010)* establish the asymptotic normality of theoretical *h*-index under a non-parametric framework. Furthermore, by applying their general results assuming two well-known distribution functions (the Pareto and Weibull distributions) for the number of citations, the authors construct confidence intervals for the empirical *h*-index of an individual researcher.

Investigations such as the above involve using mathematics and probability theory to derive statistical formulas for standard errors and confidence intervals. Often these formulas are approximations that rely on large samples. With modern computers and statistical software, resampling methods (e.g. bootstrapping) can be used to produce standard errors and confidence intervals without the use of distributional assumptions, that can prove more reliable than statistical formulas.

Among the advantages of such methodology are the requirement of fewer assumptions, since that data do not need to be normally distributed, the greater accuracy, since that the methodology does not rely on very large samples, and the generality of its use, due to that the same methodology can apply to a wide variety of statistics. In this paper we apply the non-parametric bootstrap technique for constructing confidence intervals for the *h*-index for different fields of research. In this way no specific assumptions about the distribution of the empirical *h*-index are required as well as no large samples since that the methodology is based on resampling from the initial sample. The present paper is the first work to explore the performance of bootstrap confidence intervals to estimate the *h*-index bibliometric measure. Our analysis differs to the theoretical study of *Beirlant and Einmahl (2010)* in two key points; first we do not rely on any theoretical citation distributions and second we construct field specific - and not individual – confidence intervals for the *h*-index.

The procedure for bootstrapping essentially relies on resampling from an initial sample. Specifically, one creates *B* bootstrap samples by sampling with replacement from the original data.

By applying this methodology we can estimate confidence intervals for measures of scientific performance such as the mean and median *h*-index of scientists from a given field of research, and subsequently provide means of practical value procedures, such as the selection of a superior between two candidates' scientists of the same field of research. In particular, in the remaining of the article, we introduce point estimators by implementing bootstrap methodology for the mean and median *h*-index for different fields of research using data from the highly cited researchers (HCR) database of the Institute of Scientific Information (ISI). Subsequently, confidence intervals for the true values of such estimates using four alternative bootstrap techniques are given, and the performance of the derived intervals is tested through simulations. As illustrated, the considered confidence intervals can attain, in most of the times, satisfactory coverage of the true values of the parameters.

The rest of the paper is structured as follows: In the next section we present an overview of bibliography associated with variation of bibliometric indicators for different fields of research and attempts for their normalization. Data used for our analysis are described in section 3. Bootstrap methodology implemented for the analysis is presented in section 4. Results of the analysis are presented in sections 5, 6 and 7. Finally the summary and conclusions are given in section 8.

## 2. Comparing Scientific Research for the Various Fields of Research – An Overview

An important disadvantage of the *h*-index is that typically it cannot take into account the specific field of research of a researcher. In other words, trying to compare the *h*-indices of two scientists of different fields of research is not at all a straightforward procedure, since publication rates as well as citation rates vary significantly from one field to another. As reported by *Adler et al. (2009)* the average citations per article in life sciences is about 6 times higher than in mathematics and computer sciences, making direct comparisons of citation outputs between scientists of these two disciplines invalid. *Bletsas and Sahalos (2009)* have already stressed out the need for finding a more accurate way of measuring research performance of a researcher than solely based on the single *h*-index, by utilizing other relevant measures based on the citation distribution of the researcher, such as average number of citations per publication for the particular scientific field of researcher.

In general, normalization of bibliometric indicators to account for interdisciplinary differences has already been considered in the literature (see, e.g. *van Raan 2005; Podlubny 2005; Podlubny and Kassayova 2006*). However, relatively little work has been done in this direction, in relation to the *h*-index and its modifications.

According to *Hirsch (2005),* there is considerable variation in the skewness of citation distributions even within a given subfield, and for an author with a relatively low *h* that has a few seminal papers with extraordinarily high citation counts, the *h* index will not fully reflect that scientist's accomplishments. Scientists in life sciences tend to achieve much higher *h*-values when compared to scientists in physics. For instance, in physics, a moderately productive scientist usually has an *h* equal to the number of years of service while biomedical scientists tend to have higher *h* values (*Hirsch 2005*).

As a natural consequence, there are differences in typical *h* values in different fields, determined in part by the average number of references in a paper in the field, the average number of papers produced by each scientist in the field (although, to a first approximation in a larger field, there are more scientists to share a larger number of citations, so typical *h* values should not necessarily be larger). Scientists working in nonmainstream areas will not achieve the same very high *h* values as the top echelon of those working in highly topical areas. It can be seen that, not surprisingly, all of these HCRs also have high *h* indices and that high *h* indices in the life sciences are much higher than in physics.

These latter results confirm that *h* indices in biological sciences tend to be higher than in physics. However, they also indicate that the difference appears to be much higher at the high end than on average. Clearly, more research in understanding similarities and differences of *h* index distributions in different fields of science would be of interest.

Thus, prior to comparisons of the *h*-index, in such situations some kind of "normalization" of the *h*-indices is required.

A first step in this direction is due to *Batista, Campiteli, Kinouchi and Martinez (2006)* who noticed that the number of citations a paper receives can be influenced by the number of authors. Since that the greater the number of authors, the greater the number of self-citations and the co-authorship behavior is characteristic of each discipline, thus they propose a complementary to the *h* index, $h_I$ index to quantify an individual's scientific research output valid across disciplines.

*Iglesias and Pecharromán (2007a)* [see also *Iglesias and Pecharromán 2007b*] propose a scaling of the *h*-index to account for the different scientific fields of researchers, assuming a stochastic model for the number of citations (specifically the distribution of the number of citations is assumed to be the Zipf distribution) which leads to the following expression for the theoretical *h*-index: $h = \sqrt[3]{\dfrac{N_p}{4}} \chi^{2/3}$, where $N_p$ denotes the total number of papers published and $\chi$ is the average number of citations per paper for the researcher. Based on the above specifications, *Iglesias and Pecharromán (2007a)* suggest using as a normalizing factor for the *h*-index the following expression: $f_i = \left( \chi_{physics} / \chi_i \right)^{2/3}$, where $\chi_i$ is the average number of citations per paper of scientific field i, and $\chi_{physics}$ (which is the average number of citations per paper for the Physics field) stands as the reference category. Thus, the normalized *h*-index is given by: $h_{normalized} = f_i \times h = \left( \chi_{physics} / \chi_i \right)^{2/3} \times h$.

The normalization methodology is applied to a real dataset comprising *h*-index values of HCRs affiliated with Spanish institutions. The results show that after correction with the normalizing factor, the *h* values become more homogeneous. The authors also note that correction found particularly useful in the field of mathematics, where HCRs share *h*-index values considerably lower when compared to HCRs of other disciplines.

*Radicchi, Fortunato and Castellano (2008)* study the distributions of citations received by a single publication with several disciplines, spanning broad areas of science. They show that the probability that an article is cited *c* times has large variations between different disciplines, but all distributions are rescaled on a universal curve when the relative indicator $c_f = c/c_0$ is considered, where $c_0$ is the average number of citations per article for the discipline. They consider one of the most relevant factors that may hamper a fair evaluation of scientific performance: field variation. Publications in certain disciplines are typically cited much more or much less than in others. This may happen for several reasons, including uneven number of cited papers per article in different fields or unbalanced cross-discipline citations. The authors take as normalizing factor the quantity for citations of articles belonging to a given scientific field to be the average number $c_0$ of citations received by all articles in that discipline published in the same year.

A first step toward properly taking into account field variations is to recognize that the differences in the bare citation distributions are essentially not due to specific discipline-dependent factors, but are instead related to the pattern of citations in the field, as measured by the average number of citations per article, $c_0$. It is natural then to try to factor out the bias included by the difference in the value of $c_0$ by considering a relative indicator, that is, measuring the success of a publication by the ratio $c_f = c/c_0$ between the number of citations received and the average number of citations received by articles published in its field in the same year.

By rescaling the distribution of citations for publications in a scientific discipline by their average number, a universal curve is found, independent of the specific discipline. Subsequently, they propose the use of the generalized *h*-index.

The identification of the relative indicator $c_f$ as the correct metrics to compare articles in different disciplines naturally suggests its use in a generalized version of the $h$ index, taking properly into account different citation patterns across disciplines. A crucial ingredient of the $h$ index is the number of articles published by an author. Such a quantity depends on the discipline considered. In some disciplines, the average number of articles published by an author in a year is much larger than in others. However, also in this case, this variability is rescaled away if the number $N$ of publications in a year by an author is divided by the average value in the discipline $N_0$.

The universality of this characteristic allows one to define a generalized $h$ index, $h_f$, that factors out also the additional bias due to different publication rates, thus allowing comparisons among scientists working in different fields. To compute the index for an author, his/her articles are ordered according to $c_f = c / c_0$ and this value is plotted versus the reduced rank $r / N_0$ with $r$ being the rank. In analogy with the original definition by Hirsch, the generalized index is then given by the last value of $r / N_0$ such that the corresponding $c_f$ is larger than $r / N_0$. Recently, *Waltman et al. (2012)* presented a validation study of the generalized index, and their results indicated that the scaling using the generalized index was not adequate in obtaining similar citation distributions for different fields of research.

*Namazi and Fallahzadeh (2010)* argue that calculation of the relative indicator $c_f = c / c_0$, of *Radicchi et al. (2008)* is not easy, making it a rather nominal rather than a pragmatic index. They propose the *n*-index, which is the researcher's *h*-index divided by the highest *h*-index of the journals of his/her major field of study. According to the authors, this index can surmount the problem of unequal citations in different fields and can be easily calculated. A critical study against the use of the *n*-index for correcting field variations - mainly due to that the proposed index is by its own subject to field variations since that for instance a researcher may have publications in interdisciplinary journals - can be found in *Feily and Yaghoobi (2010)*.

Finally, *Bornmann and Daniel (2009)* propose the use of z-scores instead of the relative indicator $c_f$ for normalizing citation performance across different fields of research, and by utilizing a dataset on the peer review process of AC-IE chemistry journal they test *Radicchi et al.'s (2008)* $c_f$ indicator against the well-known z-score. A similar scaling of the *h*-index to account for the field of research can be found in *Bletsas and Sahalos (2009)*.

## 3. Data

Data for the subsequent analysis were compiled from the HCRs database of the institute of scientific information (ISI). We decided to use data on HCRs of the ISI covering 21 disciplines and 6,103 researchers. These data are freely available by the Thomson Scientific (http://hcr3.isiknowledge.com/). The specific database covers the 19-year time period between 1981 and 1999 and is based on the current affiliation of the HCRs. The choice of using the specific data is driven by the fact that HCRs are a good indicator of quality research in each science field since they have a significant impact on the advancement of sciences.

For the purpose of our investigation, we have chosen to use data on the HCRs in a series of seven fields of research in total; specifically we have collected and present here related information of the HCRs in mathematics, clinical medicine, social sciences, economics, computer science, chemistry and physics. The Thomson database lists 2,307 HCRs related to the specific disciplines.

For our purposes, we have collected information associated with the *h*-index values of the HCRs in each one of the selected fields of research. The relative information was found partly in the HCR database and in the Thomson ISI Web of Science database from where the *h*-indices of the selected researchers were obtained. The *h*-indices of the researchers selected from the database were calculated up to 2012 (period of data collection: May, 2012). Specifically, approximately a 10% sample (i.e. 31 researchers in total) was identified from the HC scientists included in the database from each field, using simple random sampling.

Our motivation for choosing to investigate more than one fields of research was the belief that the patterns of the *h*-index values of the researchers would vary from one field to another.

### 4. Bootstrap Confidence Intervals

An accurate estimate of the uncertainty associated with parameter estimates is important to avoid misleading inference. This uncertainty is usually summarized by a confidence interval or region which is claimed to include the true parameter value with a specified probability. One of the most important issues in econometrics/statistics is to measure the accuracy of an estimator. The best way to do this consists in using the sampling distribution of the specific estimator. A possible way to perform this, is by the use of the bootstrap method (*Efron 1982*).

The method of non-parametric resampling using bootstrap methodology for calculating estimates and confidence intervals for the parameter of interest is based on the following general scheme:

i.      Sample n observations randomly with replacement from the initial sample of data, say vector **y_obs**.

ii.     Calculate the bootstrap version of the statistic of interest, say $\hat{\theta}$.

iii.    Repeat steps i and ii a large number of times, say $B$, to obtain an estimate of the bootstrap distribution.

This section is devoted to the construction of confidence intervals for the indices defined. As it is well-known bootstrap is a non-parametric technique that can be used whenever it is troublesome to create confidence intervals for a parameter using standard statistical techniques. A detailed description of it and its implementation for the construction of confidence intervals can be found in *Efron and Tibshirani (1993).*
 In this section we illustrate how the bootstrap method is used for constructing confidence intervals for the indices that were defined previously.

Let us assume that we have a sample of n observations. From this initial sample we generate a large number of samples, say $B$, by sampling with replacement. The choice of $B$ is arbitrary, but its value must be sufficiently large. In practice, the number of $B$ that is preferred is 1,000. The $B$ samples are called bootstrap samples. For each bootstrap sample the value of the index $h$ is calculated. After the assessment of all $B$ index values, we order them in a non-decreasing order and we denote the ith of these values by: $h_{(i)}, i = 1, ..., B$ .

In the sequel we describe the four alternative methods that we have applied in order to create bootstrap confidence intervals for the mean and median *h*-index estimates for the various fields of research using as our initial data the samples of *h*-indices collected from HCRs. Specifically, these methods are the basic bootstrap (BB), the normal bootstrap (NB), the percentile bootstrap (PB) and the bias-corrected accelerated (BCa) bootstrap.

**The Normal Bootstrap confidence interval**

According to this method, a 100(1-α)% confidence interval for the *h* index is given by:

$$\left( \hat{h} - z_{1-\alpha/2} S_h, \hat{h} + z_{1-\alpha/2} S_h \right),$$

where $z_\alpha$ denotes the 100α% percentile of the standard normal distribution,

$$S_I = \sqrt{\frac{1}{B-1} \sum_{i=1}^{B} \left( h_{(i)} - \bar{h} \right)^2}$$

is the standard deviation of the *B* index values,

$$\bar{h} = \frac{1}{B} \sum_{i=1}^{B} h_{(i)}$$

is the mean of the *B* index values and $\hat{h}$ is the index value that was assessed from the initial sample.

**The Percentile Bootstrap**

According to this approach, the 100(1-α)% confidence limits for the index *h* are the 100(α/2)% and 100(1-α/2)% percentile points of the bootstrap distribution of *h*. Consequently, the interval is:

$$\left( h_{(B\alpha/2)}, h_{(B(1-\alpha/2))} \right)$$

The PB confidence intervals may not have the correct coverage when the sampling distribution is skewed (*Davison and Hinkley 1997*). Other methods adjust the confidence interval endpoints to increase the accuracy of the coverage, such as basic or Bias-corrected.

**The Basic (non-studentized) Bootstrap**

The BB calculates endpoints by inverting hypothesis tests (*Davison and Hinkley 1997*). The upper quantile of a bootstrap distribution is used to calculate the lower confidence bound and the lower quantile is used to calculate the upper bound. When the bootstrap distribution is symmetrical about the estimate from the original data, i.e., the BB produces the same endpoints as the PB. When the distribution is skewed, the endpoints of the two methods differ.

**The Bias-corrected Accelerated (BCa) Bootstrap**

*Efron (1987)* (see also *DiCiccio and Efron 1996*) proposed a bias corrected accelerated (BCa) method which is known to improve on percentile intervals for a population parameter in traditional n>p framework. The BCa method assumes the existence of a transformation such that the parameter of interest and its estimate are transformed into a statistic having an asymptotically normal distribution.

This approach is similar to the PB but involves a slight correction. The reason why this correction is made is the potential bias. This method, despite the fact that it is more complicated than the two previously described, performs usually better than they do. According to this method, we firstly find the two successive values $h_{(i)}$ and $h_{(i+1)}$ between which the value of the index that was assessed from the initial sample $\left(\hat{h}\right)$ lies. Then, we assess the value for which the cumulative distribution function of the standard normal distribution $\Phi$ takes the value i/$B$. If we denote this value by $z_0$, then $z_0 = \Phi^{-1}(i/B)$. Finally, we calculate the probabilities $p_l$ and $p_u$ which are defined as:

$p_l = \Phi\left(2z_0 + z_{\alpha/2}\right)$ and $p_u = \Phi\left(2z_0 + z_{1-\alpha/2}\right)$.

Using these probabilities we end up with a 100(1-α)% confidence interval of the form:

$\left(h_{(B \cdot p_l)}, h_{(B \cdot pu)}\right)$.

## 5. Results of the Analysis

In the current section we present the results of the four confidence intervals, i.e. the BB, the NB, the PB and the BCa bootstrap (for 90% and 95% confidence level) for the seven disciplines under study, for estimating mean and median *h*-index values. The relationship between the sample values of the *h*-index for the seven disciplines and the estimated confidence intervals is presented and discussed.

Table A1 in the appendix presents the descriptive statistics for the 7 samples selected from the various disciplines. By a first look at the results it is evident that large variations exist between the *h*-index values of the scientists of the aforementioned fields. Lowest averages in the *h*-index are those in computer science (average *h*-index: 19.9, median *h*-index: 18), followed by mathematics category (average *h*-index: 30.84, median *h*-index: 26) and economics/business category (average *h*-index: 31, median *h*-index: 31). On the other hand, highest values are shown in the clinical medicine (average *h*-index: 95.87, median *h*-index: 94) and in chemistry (average *h*-index: 81.52, median *h*-index: 77). The average *h*-index from the total sample, independently of the discipline is 52.02, whereas the median from total sample was 37.

From our initial sample, a number of *B*=1,000 bootstrap samples were generated in order to apply our methodology.

Table A2 in the appendix presents the average, median and standard error of the 1,000 bootstrap estimates of *h*-index. Figures A1 and A2 in Appendix show the distribution of the mean and median *h*-index values for each discipline, respectively.

Next, a 95% and a 90% bootstrap confidence interval has been constructed by each of the four methods, i.e. NB, BB, PB and BCa for the mean and median *h*-index for the seven disciplines (see Tables A3 and A4). For instance, the 95% PB confidence interval for the mean *h*-index in the discipline of mathematics is [26.32, 36.19], whereas the corresponding 90% PB confidence interval produces a shorter average length, given by [26.87, 35.26]. Fields with the most similar confidence intervals are mathematics and economics/business, with the field of economics presenting narrower lengths. Field of research with the lowest CIs is that of computer science. On the opposite side, fields of clinical medicine, chemistry and physics are the fields with the highest intervals.

The most astonishing outcome derived by a first inspection of the constructed CIs is that for the majority of intervals, both for mean and median $h$-index, there is no overlap between them. This of course is a clear indication that the $h$-index values of scientists of distinct different fields of research are not at all comparable to each other, with most distinct examples being for instance the fields of clinical medicine (highest CI) and computer science (lowest CI). It is thus evident, that direct comparisons can be absolutely misleading even for any other two fields of research – except only for scientists of the fields of mathematics and economics/business. This is mostly obvious for the CIs of the mean $h$-index, however even for the CIs for the median $h$-index the overlaps between CIs for the different disciplines are minor.

Generally, as concerns the mean $h$-index, method BCa gives the wider intervals in most of the cases, while NB gives shorter ones. BB and PB methods produce almost identical intervals as regards their ranges. Wider CIs are obtained for chemistry, followed by clinical medicine, social sciences, physics, mathematics, economics/business and finally the narrower CIs are shown for computer science field.

As concerns the intervals of the median $h$-index, NB gives the wider intervals in most of the cases. Wider CIs are obtained for chemistry and social sciences, followed by computer science, physics, clinical medicine and economics. The narrower CIs are shown for mathematics field.

For illustration purposes, the 90% BB confidence intervals for $B$=1,000 bootstrap replications for the parameter of the average $h$-index are presented in Figure 1.

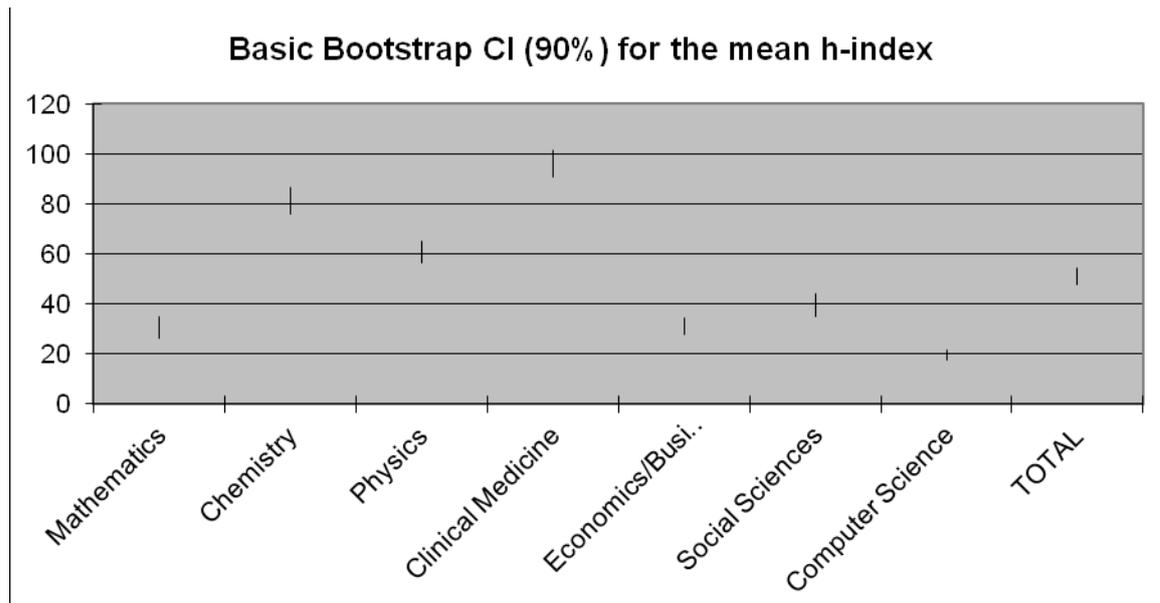

**Figure 1:** 90% Basic bootstrap confidence intervals for the average *h*-index for the seven disciplines

## 6. Observed Coverage of the CIs

In order to illustrate the assessment of the *h*-index and the construction of confidence intervals for its true values we used the data on HCRs to find the observed coverage of the four methods for constructing the confidence intervals. In each field of research we found the observed coverage, which must be as close as possible to the nominal coverage (i.e. 0.9 and 0.95, respectively). Tables 1 and 2 present observed coverage (OC) (i.e. the proportion of times where the value of the parameter falls within the interval) along with the lower miss rates (i.e. the proportion of times where the value of the parameter is smaller than the lower limit of the interval) and upper miss rates (i.e. the proportion of times where the value of the parameter is larger than the upper limit of the interval) that each method gives. As illustrated, the bootstrap confidence intervals can attain, in most of the times, quite satisfactory coverage, hence we may argue that they can be implemented each time one wishes to make comparisons of scientists from a specific field of research.

The field of research with the worst coverage as concerns the mean *h*-index appears to be computer science. Its less adequate coverage may be attributed to the fact that the specific field presents the most noticeable variations as concerns the *h*- indices of the related HCRs. (However, the coverage of the median *h*-index of the computer science field is substantially improved for PB and BCa confidence intervals, whereas an extremely poor coverage is observed for BB and NB intervals suggesting that the former intervals probably should be preferred for constructing confidence intervals in the specific field). As the values of upper miss rates for BB and NB reveal for the intervals of median *h*-indices in computer science, the two intervals underestimate the parameters, failing to include the values in the right long tail of their distribution.

Generally, all fields attain satisfactory to excellent coverage in most of the cases and for most of the bootstrap confidence interval methods, with exceptions though. More specifically, methods that attains most satisfactory performance (i.e. best coverage[1]) for the mean *h*-index as concerns the 90% confidence level are PB and BCa, whereas the best coverage for the 95% level is attained by NB method[2].

When we examine coverage of the median *h*-index we find that now PB method outperforms the other three, since that it exceeds the best coverage for both the 90% and 95% confidence levels. Specifically, PB has shown best coverage of median *h*-index in mathematics, chemistry, physics, economics, social sciences and medicine for the 90% level, and mathematics, chemistry, physics, economics, medicine and computer science for the 95% level. BB, NB and BCa seem to fail to attain adequate CIs in the case of the median *h*-index in several occasions.

In summary, among the four alternative procedures for assessing bootstrap confidence intervals for the mean *h*-index we propose the use of the PB and the BCa bootstrap, whereas for assessing confidence intervals for the median *h*-index we propose using the PB. The extremely poor coverage of the mean *h*-index CI in the field of computer science may be tempting to lead us proposing the median *h*-index CIs for comparisons of scientists from the same field of research.

---

[1] PB and BCa have the best coverage in three science fields (i.e. PB in mathematics, chemistry and medicine and BCa in physics, economics and social sciences).

[2] NB has the best coverage in five science fields (i.e. mathematics, economics/business, physics, social sciences and computer science).

**Table 1:** Observed coverage (OC) and miss rates of 90% NB, PB, BB and BCa bootstrap confidence intervals for average and median *h*-index from the various fields of research

| Science Field | Bootstrap Method | OC (Mean) | Miss Rate | | OC (Median) | Miss Rate | |
|---|---|---|---|---|---|---|---|
| | | | Lower | Upper | | Lower | Upper |
| Mathematics | **BB** | 90.3 | 2.9 | 6.8 | 89.3 | 0.5 | 10.2 |
| | **PB** | 91.8 | 4.5 | 3.7 | 90.8 | 4 | 5.2 |
| | **BCa** | 91.5 | 6.2 | 2.3 | 89.6 | 0.2 | 10.2 |
| | **NB** | 90.6 | 3.4 | 6 | 89.3 | 0.5 | 10.2 |
| Chemistry | **BB** | 91 | 4.6 | 4.4 | 90.9 | 4.8 | 4.3 |
| | **PB** | 91.3 | 5.3 | 3.4 | 93.7 | 2 | 4.3 |
| | **BCa** | 91.2 | 5.2 | 3.6 | 88.4 | 2 | 9.6 |
| | **NB** | 91.2 | 4.9 | 3.9 | 93.7 | 2 | 4.3 |
| Physics | **BB** | 92.3 | 4 | 3.7 | 90.6 | 0.2 | 9.2 |
| | **PB** | 91.9 | 3.5 | 4.6 | 94 | 4.1 | 1.9 |
| | **BCa** | 92.8 | 1.4 | 5.8 | 83.4 | 0.5 | 16.1 |
| | **NB** | 92.2 | 3.6 | 4.2 | 86.7 | 4.1 | 9.2 |
| Clinical medicine | **BB** | 90.7 | 5.5 | 3.8 | 93.9 | 1.5 | 4.6 |
| | **PB** | 91.3 | 3.9 | 4.8 | 93.9 | 1.5 | 4.6 |
| | **BCa** | 90.8 | 4.7 | 4.5 | 83.1 | 0 | 16.9 |
| | **NB** | 90 | 5.5 | 4.5 | 93.9 | 1.5 | 4.6 |
| Economics/Business | **BB** | 89.1 | 6.1 | 4.8 | 80.9 | 2 | 17.1 |
| | **PB** | 89 | 5.5 | 5.5 | 93.2 | 4.9 | 1.9 |
| | **BCa** | 89.7 | 7.3 | 3 | 82.6 | 0.3 | 17.1 |
| | **NB** | 89.4 | 5.9 | 4.7 | 78 | 4.9 | 17.1 |
| Social sciences | **BB** | 89.9 | 3.4 | 6.7 | 97.7 | 0.4 | 1.9 |
| | **PB** | 89.5 | 5 | 5.5 | 97.7 | 0.4 | 1.9 |
| | **BCa** | 90.7 | 5.7 | 3.6 | 90.1 | 0.1 | 9.8 |
| | **NB** | 90.3 | 3.7 | 6 | 85.9 | 9.3 | 4.8 |
| Computer science | **BB** | 90 | 4.2 | 5.8 | 59 | 0 | 41 |

| | Bootstrap Method | OC (Mean) | Lower | Upper | OC (Median) | Lower | Upper |
|---|---|---|---|---|---|---|---|
| | **PB** | 88.7 | 6.9 | 4.4 | 96.1 | 2.5 | 1.4 |
| | **BCa** | 88.7 | 7.4 | 3.9 | 98 | 0.6 | 1.4 |
| | **NB** | 88.4 | 6 | 5.6 | 59 | 0 | 41 |
| **TOTAL** | **BB** | **91.7** | **3.3** | **5** | **77.4** | **22.5** | **0.1** |
| | **PB** | **91** | **5.7** | **3.3** | **95.1** | **1.9** | **3** |
| | **BCa** | **91** | **6** | **3** | **95.1** | **1.9** | **3** |
| | **NB** | **91** | **4.5** | **4.5** | **85.6** | **14.3** | **0.1** |

**Table 2:** Observed coverage (OC) and miss rates of 95% NB, PB, BB and BCa bootstrap confidence intervals for average and median *h*-index from the various fields of research

| Science Field | Bootstrap Method | OC (Mean) | Miss Rate | | OC (Median) | Miss Rate | |
|---|---|---|---|---|---|---|---|
| | | | Lower | Upper | | Lower | Upper |
| Mathematics | **BB** | 95.9 | 0.8 | 3.3 | 94.3 | 0.5 | 5.2 |
| | **PB** | 95.6 | 2.7 | 1.7 | 99.4 | 0.5 | 0.1 |
| | **BCa** | 95.1 | 3.9 | 1 | 89.6 | 0.2 | 10.2 |
| | **NB** | 96.4 | 1.3 | 2.3 | 94.3 | 0.5 | 5.2 |
| Chemistry | **BB** | 96.1 | 2.3 | 1.6 | 95.7 | 0 | 4.3 |
| | **PB** | 96.3 | 2.5 | 1.2 | 97.4 | 2 | 0.6 |
| | **BCa** | 96.5 | 2.5 | 1 | 94.8 | 0.9 | 4.3 |
| | **NB** | 96.2 | 2.4 | 1.4 | 95.5 | 0.2 | 4.3 |
| Physics | **BB** | 95.5 | 2.3 | 2.2 | 95.4 | 0.2 | 4.4 |
| | **PB** | 96.3 | 0.7 | 3 | 97.6 | 0.5 | 1.9 |
| | **BCa** | 95.8 | 0.5 | 3.7 | 83.7 | 0.2 | 16.1 |
| | **NB** | 96.6 | 1 | 2.4 | 95.1 | 0.5 | 4.4 |
| Clinical medicine | **BB** | 94.9 | 2.6 | 2.5 | 95.4 | 0 | 4.6 |
| | **PB** | 95.7 | 1.6 | 2.7 | 96.6 | 1.5 | 1.9 |
| | **BCa** | 95.2 | 2.1 | 2.7 | 83.1 | 0 | 16.9 |
| | NB | 94.9 | 2.5 | 2.6 | 95.1 | 0.3 | 4.6 |
| Economics/Business | BB | 94.8 | 2.4 | 2.8 | 96.1 | 2 | 1.9 |
| | PB | 94.8 | 2.4 | 2.8 | 96.1 | 2 | 1.9 |
| | BCa | 94 | 4.3 | 1.7 | 82.6 | 0.3 | 17.1 |
| | NB | 95.5 | 2.4 | 2.1 | 96.1 | 2 | 1.9 |
| Social sciences | BB | 94.5 | 1.5 | 4 | 98.1 | 0 | 1.9 |
| | PB | 94.2 | 3.1 | 2.7 | 97.7 | 0.4 | 1.9 |
| | BCa | 94 | 3.7 | 2.3 | 90.2 | 0 | 9.8 |
| | NB | 95.2 | 1.8 | 3 | 97.7 | 0.4 | 1.9 |
| Computer science | BB | 94.8 | 2.2 | 3 | 59 | 0 | 41 |
| | PB | 94.8 | 3.2 | 2 | 98 | 0.6 | 1.4 |
| | BCa | 94.7 | 3.5 | 1.8 | 98 | 0.6 | 1.4 |
| | NB | 94.9 | 2.6 | 2.5 | 90 | 0 | 10 |
| **TOTAL** | **BB** | **95.8** | **2.2** | **2** | **85.6** | **14.3** | **0.1** |
| | **PB** | **95.8** | **2.2** | **2** | **97.6** | **1.9** | **0.5** |
| | **BCa** | **95.5** | **2.6** | **1.9** | **98.7** | **0.8** | **0.5** |
| | **NB** | **95.7** | **2.2** | **2.1** | **95.1** | **4.9** | **0** |

## 7. Examining the Performance of Normalization Indices through Bootstrap CIs

In this section we attempt a test performance of some of the normalizing indices proposed in the literature for accounting for field variations, using the CIs derived from our analysis. Specifically, the normalizing *h*-index of *Iglesias and Pecharromán (2007a)* and the *n*-index of *Namazi and Fallahzadeh (2010)* are utilized to construct the corresponding CIs and then check adequacy of the latter measures in obtaining similar citation distributions for the different fields of research.

To this end, Tables A5 to A8 in the Appendix present 90% and 95% CIs for the mean/median $h_{normalizing}$ index and *n*-index respectively, calculated using bootstrap methodology (*B*=1,000). Further, figures 2 and 3 show the 90% basic bootstrap CIs for the mean $h_{normalized}$ index and the $n$-index, respectively, for visually inspecting the possible coverage between fields of research.

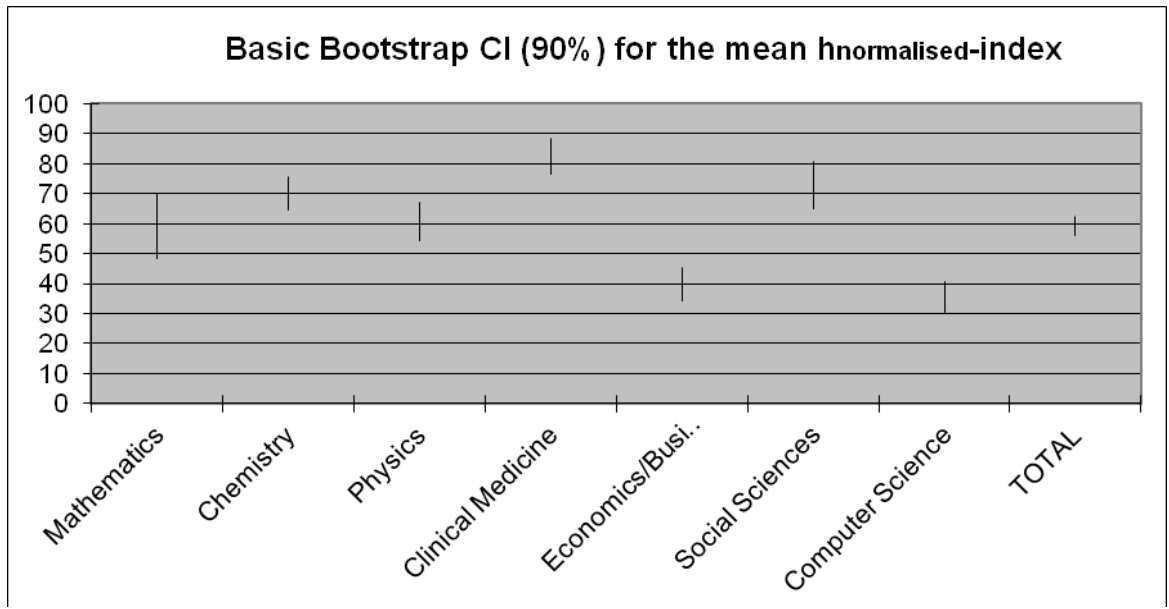

**Figure 2:** 90% Basic bootstrap confidence intervals for the average $h_{normalized}$-index for the seven disciplines

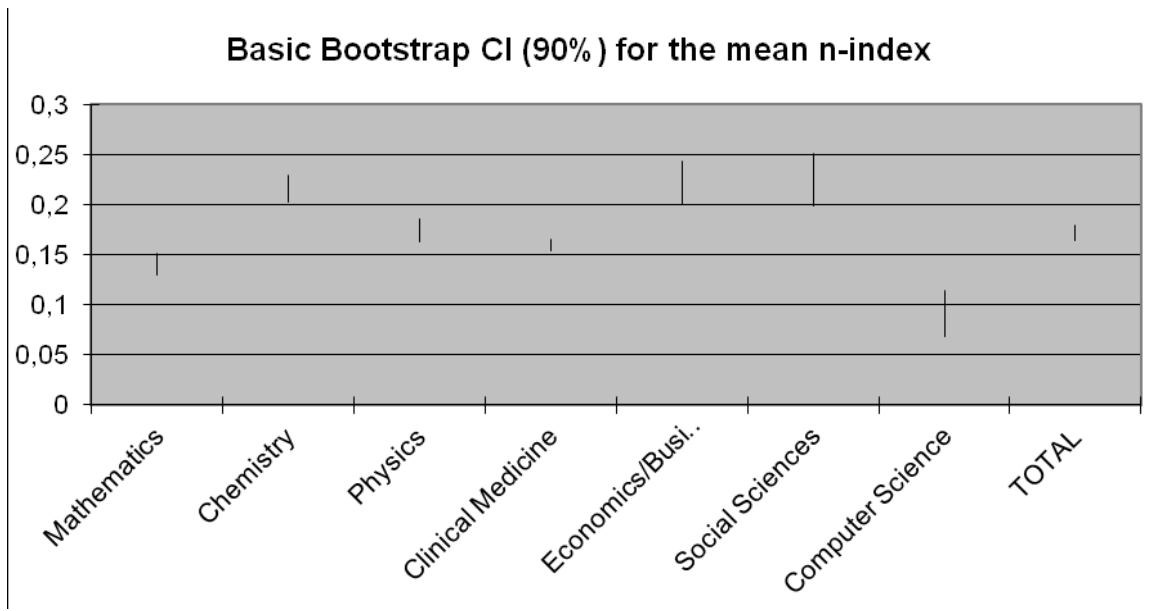

**Figure 3:** 90% Basic bootstrap confidence intervals for the average $n$-index for the seven disciplines

As one observes, both scaling of *Iglesias and Pecharromán (2007a)* and *Namazi and Fallahzadeh (2010)* fail to provide a very good collapse of the confidence intervals. Differences between the different fields remain even after both normalizations applied to the data. However, both figures present a better approximation between the fields of research when compared to the raw data of figure 1. Overlaps now, between the CIs are 7 and 6 for the $h_{normalizing}$ index and $n$-index respectively, indicating the improvement in their homogeneity. Between the two normalizations, the $n$-index seems to perform better, as the graph reveals, since that fields of chemistry, economics and social sciences are now comparable, as this also merely holds for mathematics, physics and clinical medicine.

## 8. Conclusions

The discussion given above described methods for obtaining point estimates and corresponding confidence intervals for mean and median $h$-index, in order to compare different fields of research. In doing this, an efficient strategy was proposed for the identification of confidence intervals for measuring the uncertainty of the differences between the $h$-indices of scientists belonging to different fields of research. In particular, in the current paper the non-parametric bootstrap for constructing confidence intervals for the $h$-index of scientists from different fields of research has been applied, by utilizing data on HCRs obtained from the Institute of Scientific Information (ISI). The analysis showed that no direct comparisons between scientists of different fields can be valid, except maybe for the fields of mathematics and economics. The higher intervals were observed for the clinical medicine, chemistry and physics fields. At the other extreme, computer science exhibits the lower CIs.

The wider intervals are shown in the fields of chemistry and social sciences, followed by social sciences, mathematics, physics, economics/business, whereas the narrower intervals are shown in the mathematics science field of research.

Analysis results showed that the four bootstrap methods for constructing confidence limits performed differently as regards both the field of research and the parameter estimated (mean and median $h$-index).

A combined application of both normalizing indices with construction of associated CIs may lead to a more valid and fair comparison between the $h$-indices of scientists belonging to different fields of research, than just comparing point estimates based on the normalizing procedures alone.

The suggested approach on one hand adds in the need for finding methodology to assist in the examination of field variation as concerns scientific performance and on the other hand may be proven very useful in cases one wants to compare research performance of scientists in the same discipline. Further research associated with construction of confidence intervals of other related bibliometric indices may add to the results of the current study allowing us a more comprehensive insight into the issue of field variations in research performance.

**Acknowledgements**

We would like to thank Professor M. Schreiber for his constructive suggestions and insightful comments.

**APPENDIX**

**Table A1:** Summary statistics of the *h*-index samples for the various fields of research

| Science Field | Mean | Median | Standard Deviation | Minimum | Maximum | Count |
|---|---|---|---|---|---|---|
| Mathematics | 30.84 | 26 | 14.22 | 14 | 67 | 31 |
| Chemistry | 81.52 | 77 | 18.8 | 54 | 124 | 31 |
| Physics | 60.71 | 61 | 13.8 | 22 | 91 | 31 |
| Clinical medicine | 95.87 | 94 | 17.6 | 61 | 141 | 31 |
| Economics/Business | 31 | 31 | 11.65 | 8 | 61 | 31 |
| Social sciences | 40.03 | 37 | 16.47 | 19 | 78 | 31 |
| Computer science | 19.9 | 18 | 7.53 | 7 | 43 | 31 |
| **TOTAL** | 52.02 | 37 | 33.3 | 8 | 141 | 217 |

**Table A2:** Bootstrap estimates of mean and median *h*-index (*B*=1,000)

| Science Field | mean | Bias[*] | Std. error | median | bias | Std. error |
|---|---|---|---|---|---|---|
| Mathematics | 30.84 | -0.076 | 2.524 | 26 | 0.165 | 1.278 |
| Chemistry | 81.52 | 0.072 | 3.426 | 77 | 0.093 | 3.931 |
| Physics | 60.7 | 0.273 | 2.428 | 61 | 0.036 | 2.378 |
| Clinical medicine | 95.87 | -0.037 | 3 | 94 | 0.224 | 2.064 |
| Economics/Business | 31 | 0.014 | 2.02 | 31 | 0.14 | 1.887 |
| Social sciences | 40.03 | 0.179 | 2.85 | 37 | 0.191 | 3.253 |
| Computer science | 19.9 | -0.024 | 1.329 | 18 | 1.376 | 2.855 |
| **TOTAL** | **51.41** | **-0.049** | **2.094** | **47** | **-1.165** | **5.167** |

*(\*) Bias: A measure of the systematic asymmetry between the bootstrap distribution and the sample*

**Table A3:** Bootstrap confidence intervals for the mean *h*-index (*B*=1,000)

| Science Field | NB (90%) | BB (90%) | PB (90%) | BCa (90%) | NB (95%) | BB (95%) | PB (95%) | BCa (95%) |
|---|---|---|---|---|---|---|---|---|
| Mathematics | 26.65-34.91 | 26.42-34.81 | 26.87-35.26 | 27.13-35.68 | 25.86-35.7 | 25.49-35.35 | 26.32-36.19 | 26.74-36.63 |
| Chemistry | 76.03-86.94 | 75.91-86.84 | 76.2-87.12 | 76.17-87.05 | 74.98-87.98 | 74.91-87.9 | 75.13-88.13 | 75.13-88.23 |
| Physics | 56.43-64.82 | 56.68-65.03 | 56.39-64.74 | 55.71-64.44 | 55.63-65.62 | 55.94-66.03 | 55.39-65.48 | 54.82-65 |
| Clinical medicine | 90.81-101.08 | 90.81-101.45 | 90.29-100.93 | 90.5-101.03 | 89.83-102.06 | 89.84-102.32 | 89.42-101.9 | 89.55-101.92 |
| Economics/Business | 27.79-34.43 | 27.81-34.35 | 27.65-34.19 | 28.03-34.71 | 27.16-35.07 | 27.16-34.87 | 27.13-34.84 | 27.51-35.45 |
| Social sciences | 35.26-44.57 | 35.19-44.41 | 35.65-44.87 | 35.81-45.2 | 34.37-45.46 | 34.13-45.16 | 34.91-45.93 | 35.26-46.23 |
| Computer science | 17.75-21.99 | 17.58-21.94 | 17.87-22.22 | 17.93-22.26 | 17.34-22.4 | 17.23-22.35 | 17.45-22.58 | 17.51-22.61 |
| **TOTAL** | **47.92-54.82** | **47.7-54.65** | **48.17-55.12** | **48.26-55.16** | **47.26-55.48** | **47.23-55.58** | **47.24-55.59** | **47.37-55.71** |

NB: Normal Bootstrap
BB: Basic Bootstrap
PB: Percentile Bootstrap
Bca: Bias-corrected adjusted Bootstrap

**Table A4:** Bootstrap confidence intervals for the median *h*-index (*B*=1,000)

| Science Field | NB (90%) | BB (90%) | PB (90%) | BCa (90%) | NB (95%) | BB (95%) | PB (95%) | BCa (95%) |
|---|---|---|---|---|---|---|---|---|
| Mathematics | 23.8-27.76 | 24-27 | 25-28 | 22-27 | 23.42-28.14 | 23-28 | 24-29 | 22-27 |
| Chemistry | 70.2-83.44 | 72-83 | 71-82 | 71-79 | 68.93-84.71 | 62-83 | 71-92 | 70-82 |
| Physics | 57.19-64.73 | 56-64 | 58-66 | 56.33-62 | 56.47-65.45 | 56-65 | 57-66 | 53.85-62 |
| Clinical medicine | 90.29-97.32 | 91-97 | 91-97 | 89-95 | 89.62-97.99 | 89-97 | 91-99 | 87-95 |
| Economics/Business | 28.01-33.74 | 28-33 | 29-34 | 24-32 | 27.46-34.29 | 28-34 | 28-34 | 23.62-32 |
| Social sciences | 31.43-42.66 | 31-43 | 31-43 | 29-39 | 30.35-43.74 | 25.15-43 | 31-48.85 | 27.08-39 |
| Computer science | 11.95-21.32 | 12-20 | 16-24 | 15-24 | 11.05-22.22 | 12-21 | 15-24 | 15-24 |
| **TOTAL** | **39.36-57.07** | **41-57** | **37-53** | **37-53** | **37.66-58.77** | **40-57** | **37-54** | **35-54** |

NB: Normal Bootstrap
BB: Basic Bootstrap
PB: Percentile Bootstrap
Bca: Bias-corrected adjusted Bootstrap

**Table A5:** Bootstrap confidence intervals for the mean $h_{normalized}$-index ($B$=1,000)

| Science Field | NB (90%) | BB (90%) | PB (90%) | BCa (90%) | NB (95%) | BB (95%) | PB (95%) | BCa (95%) |
|---|---|---|---|---|---|---|---|---|
| Mathematics | 49.01-64.23 | 48.76-63.7 | 49.17-64.11 | 49.94-66.02 | 47.55-65.68 | 46.81-64.82 | 48.05-66.06 | 49.06-67.2 |
| Chemistry | 70.02-79.79 | 70.1-79.77 | 70.22-79.89 | 70.21-79.89 | 69.08-80.72 | 69.33-80.72 | 69.27-80.66 | 69.27-80.69 |
| Physics | 56.67-64.5 | 56.75-64.48 | 56.94-64.67 | 56.46-64.26 | 55.92-65.25 | 56.1-65.45 | 55.97-65.32 | 55.51-64.94 |
| Clinical medicine | 69.12-76.75 | 68.89-76.76 | 68.97-76.83 | 69.24-77.15 | 68.39-77.48 | 68.13-77.54 | 68.18-77.59 | 68.47-77.87 |
| Economics/Business | 36.65-45.21 | 36.7-45.09 | 36.75-45.14 | 36.96-45.43 | 35.83-46.03 | 35.98-46.07 | 35.77-45.86 | 35.99-45.96 |
| Social sciences | 56.15-71.38 | 56.26-71.07 | 57.04-71.85 | 57.1-71.85 | 54.69-72.84 | 54.41-72.52 | 55.59-73.7 | 55.68-73.75 |
| Computer science | 31.22-38.69 | 31.33-38.84 | 31.05-38.44 | 31.33-38.84 | 30.51-39.4 | 30.43-39.18 | 30.49-39.23 | 30.88-39.71 |
| **TOTAL** | **55.2-60.37** | **55.14-60.26** | **55.4-60.52** | **55.35-60.5** | **54.71-60.87** | **54.78-60.77** | **54.89-60.88** | **54.83-60.86** |

NB: Normal Bootstrap
BB: Basic Bootstrap
PB: Percentile Bootstrap
Bca: Bias-corrected adjusted Bootstrap

**Table A6:** Bootstrap confidence intervals for the median $h_{normalized}$-index ($B$=1,000)

| Science Field | NB (90%) | BB (90%) | PB (90%) | BCa (90%) | NB (95%) | BB (95%) | PB (95%) | BCa (95%) |
|---|---|---|---|---|---|---|---|---|
| Mathematics | 43.45-50.93 | 43.92-49.41 | 45.75-51.24 | 40.26-49.41 | 42.73-51.64 | 42.09-51.24 | 43.92-53.07 | 38.43-49.41 |
| Chemistry | 64.8-76.33 | 58.88-75.44 | 66.24-82.8 | 65.32-75.44 | 63.69-77.44 | 58.88-76.36 | 65.32-82.8 | 64.4-82.8 |
| Physics | 57.13-64.87 | 57-64 | 58-65 | 56-62 | 56.39-65.62 | 56-65 | 57-66 | 53-63 |
| Clinical medicine | 69.02-73.81 | 69.16-72.96 | 69.92-73.72 | 67.64-72.2 | 68.56-74.27 | 67.64-73.72 | 69.16-75.24 | 67.64-72.2 |
| Economics/Business | 36.49-44.76 | 36.96-43.56 | 38.28-44.88 | 30.36-40.92 | 35.7-45.55 | 36.96-44.88 | 36.96-44.88 | 30.36-42.24 |
| Social sciences | 51.13-67.86 | 54.4-68.8 | 49.6-64 | 46.4-62.4 | 49.53-69.46 | 49.6-68.8 | 49.6-68.8 | 46.4-62.4 |
| Computer science | 21.19-37.32 | 21-35 | 28-42 | 27.16-38.5 | 19.65-38.86 | 21-35 | 28-42 | 26.25-42 |
| **TOTAL** | **51.45-60.29** | **50.49-60.13** | **50.75-60.39** | **50.75-60.23** | **50.6-61.14** | **50.16-60.72** | **50.16-60.72** | **50.16-60.72** |

NB: Normal Bootstrap
BB: Basic Bootstrap
PB: Percentile Bootstrap
Bca: Bias-corrected adjusted Bootstrap

**Table A7:** Bootstrap confidence intervals for the mean *n*-index (*B*=1,000)

| Science Field | NB (90%) | BB (90%) | PB (90%) | BCa (90%) | NB (95%) | BB (95%) | PB (95%) | BCa (95%) |
|---|---|---|---|---|---|---|---|---|
| Mathematics | 0.144-0.189 | 0.143-0.188 | 0.145-0.19 | 0.145-0.189 | 0.139-0.193 | 0.137-0.192 | 0.141-0.197 | 0.141-0.197 |
| Chemistry | 0.213-0.244 | 0.213-0.244 | 0.214-0.245 | 0.215-0.245 | 0.21-0.247 | 0.21-0.248 | 0.21-0.247 | 0.211-0.249 |
| Physics | 0.162-0.186 | 0.162-0.185 | 0.163-0.186 | 0.163-0.186 | 0.16-0.188 | 0.16-0.189 | 0.159-0.188 | 0.16-0.188 |
| Clinical medicine | 0.154-0.171 | 0.153-0.17 | 0.155-0.172 | 0.154-0.172 | 0.152-0.173 | 0.151-0.173 | 0.152-0.174 | 0.152-0.173 |
| Economics/Business | 0.197-0.246 | 0.198-0.246 | 0.197-0.245 | 0.199-0.248 | 0.192-0.251 | 0.19-0.252 | 0.191-0.252 | 0.193-0.255 |
| Social sciences | 0.205-0.262 | 0.204-0.261 | 0.207-0.264 | 0.209-0.265 | 0.2-0.268 | 0.199-0.266 | 0.202-0.269 | 0.204-0.272 |
| Computer science | 0.1-0.126 | 0.099-0.125 | 0.1-0.126 | 0.101-0.127 | 0.098-0.128 | 0.098-0.128 | 0.098-0.128 | 0.099-0.129 |
| **TOTAL** | **0.177-0.194** | **0.178-0.194** | **0.177-0.194** | **0.177-0.194** | **0.176-0.196** | **0.176-0.196** | **0.176-0.196** | **0.176-0.196** |

**Table A8:** Bootstrap confidence intervals for the median *n*-index (*B*=1,000)

| Science Field | NB (90%) | BB (90%) | PB (90%) | BCa (90%) | NB (95%) | BB (95%) | PB (95%) | BCa (95%) |
|---|---|---|---|---|---|---|---|---|
| Mathematics | 0.128-0.15 | 0.13-0.151 | 0.13-0.151 | 0.119-0.146 | 0.126-0.152 | 0.124-0.151 | 0.13-0.157 | 0.117-0.146 |
| Chemistry | 0.198-0.234 | 0.202-0.23 | 0.202-0.23 | 0.199-0.23 | 0.194-0.237 | 0.18-0.233 | 0.18-0.253 | 0.197-0.253 |
| Physics | 0.164-0.186 | 0.163-0.186 | 0.163-0.186 | 0.154-0.178 | 0.161-0.188 | 0.16-0.186 | 0.163-0.189 | 0.152-0.178 |
| Clinical medicine | 0.153-0.165 | 0.154-0.165 | 0.153-0.165 | 0.151-0.161 | 0.152-0.166 | 0.151-0.165 | 0.154-0.168 | 0.151-0.163 |
| Economics/Business | 0.198-0.243 | 0.2-0.243 | 0.2-0.243 | 0.171-0.229 | 0.193-0.247 | 0.2-0.243 | 0.2-0.243 | 0.17-0.243 |
| Social sciences | 0.187-0.246 | 0.199-0.251 | 0.181-0.234 | 0.17-0.228 | 0.181-0.252 | 0.181-0.251 | 0.181-0.251 | 0.17-0.228 |
| Computer science | 0.067-0.12 | 0.068-0.114 | 0.09-0.136 | 0.085-0.125 | 0.062-0.125 | 0.062-0.114 | 0.09-0.142 | 0.085-0.136 |
| **TOTAL** | **0.163-0.182** | **0.164-0.18** | **0.163-0.179** | **0.161-0.179** | **0.161-0.183** | **0.162-0.182** | **0.161-0.181** | **0.16-0.181** |

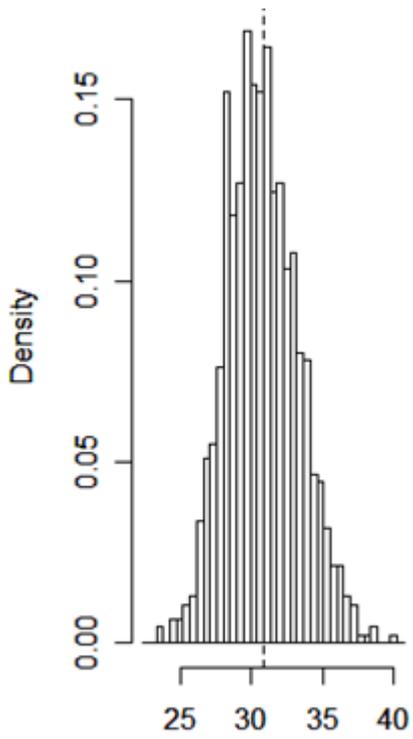

Mean *h*-index (Mathematics)

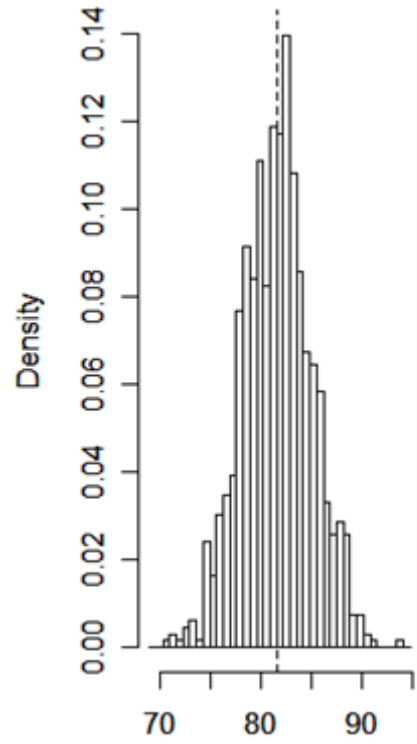

Mean *h*-index (Chemistry)

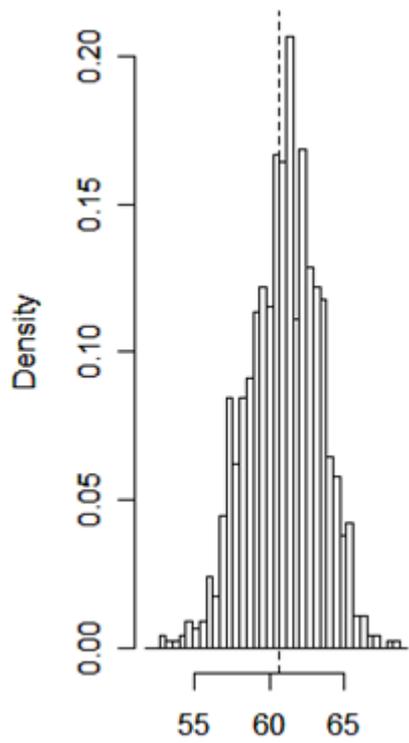

Mean *h*-index (Physics)

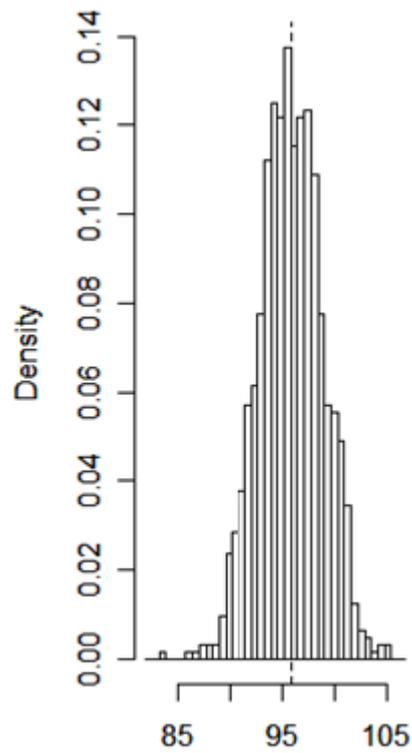

Mean *h*-index (Clinical medicine)

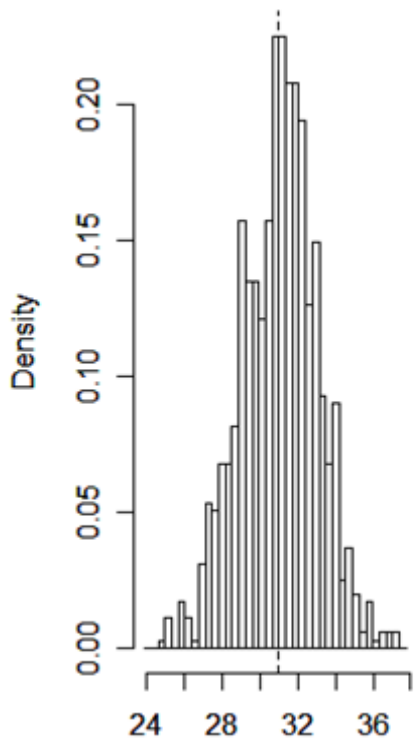

Mean *h*-index (Economics)

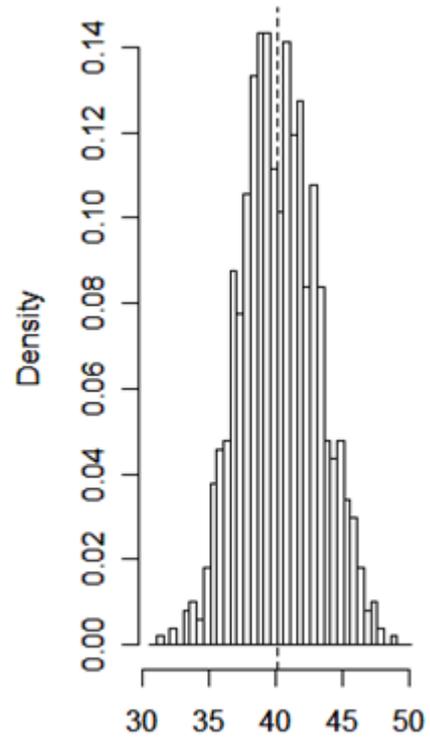

Mean *h*-index (Social sciences)

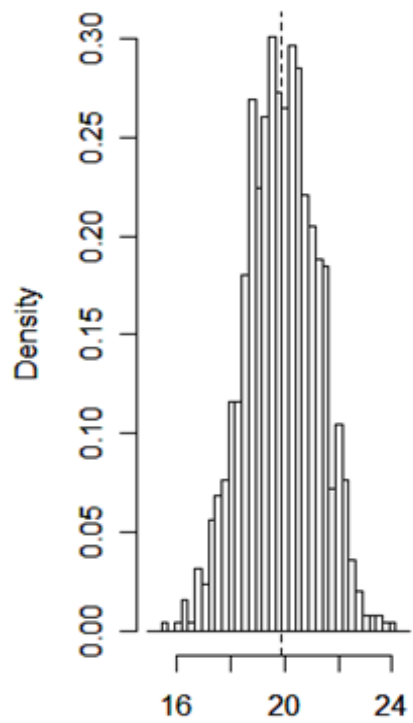

Mean *h*-index (Computer science)

**Figure A1: Bootstrap distributions for the mean *h*-index**

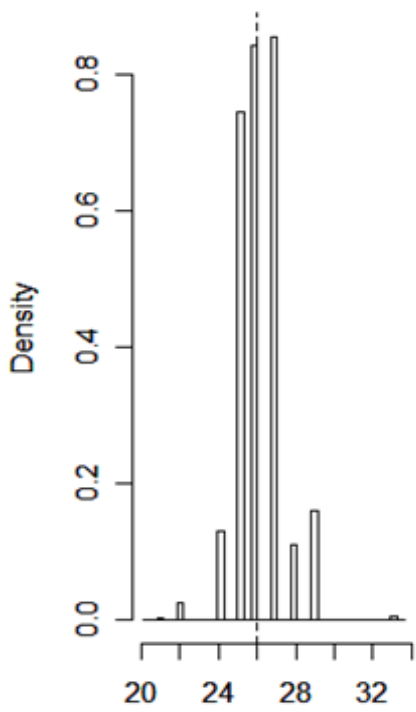

Median *h*-index (Mathematics)

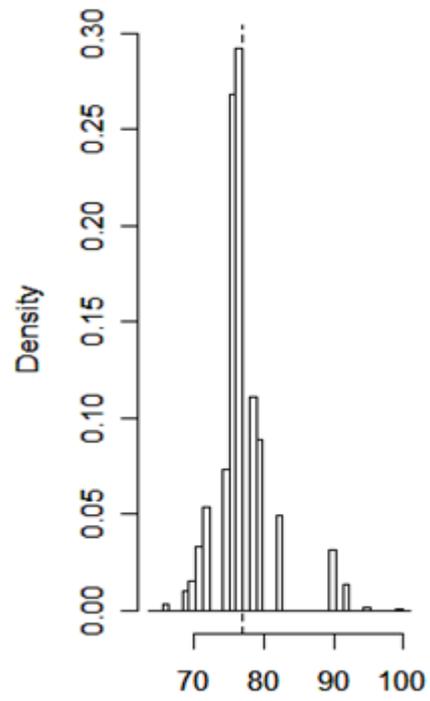

Median *h*-index (Chemistry)

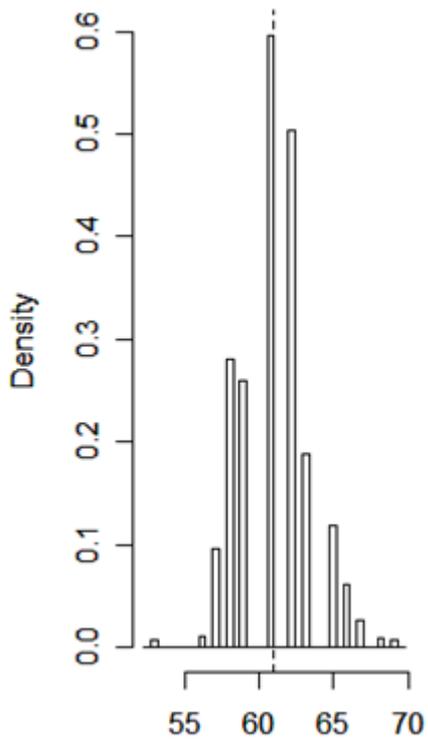

Median *h*-index (Physics)

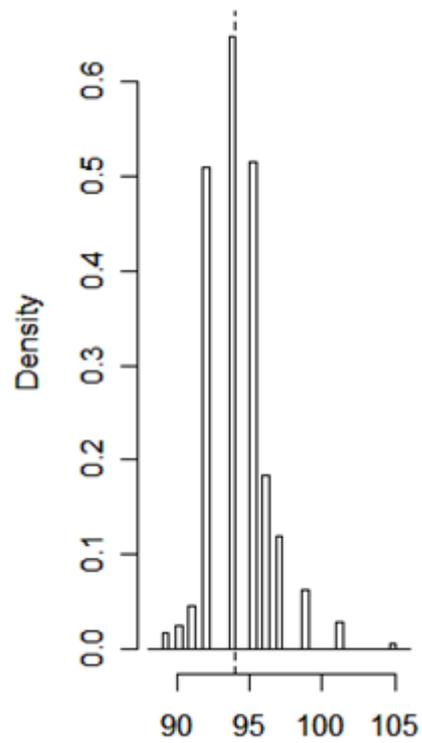

Median *h*-index (Clinical medicine)

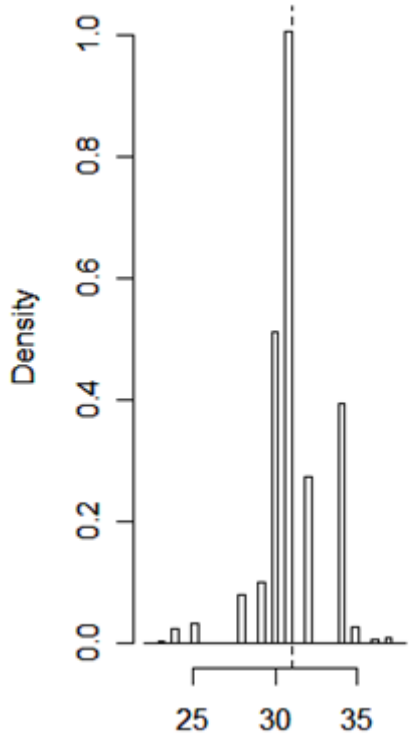
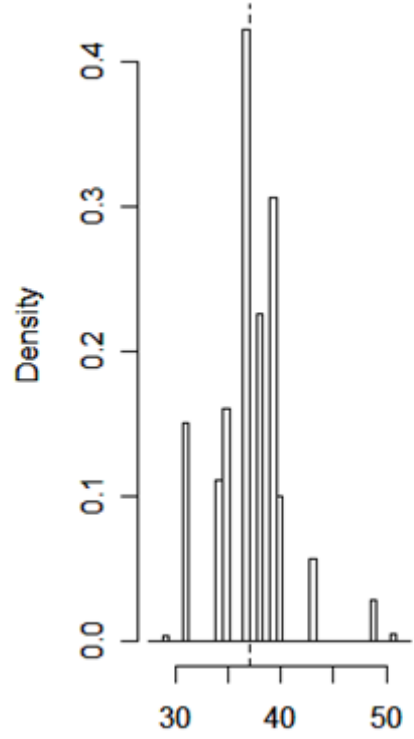

Median *h*-index (Economics)          Median *h*-index (Social sciences)

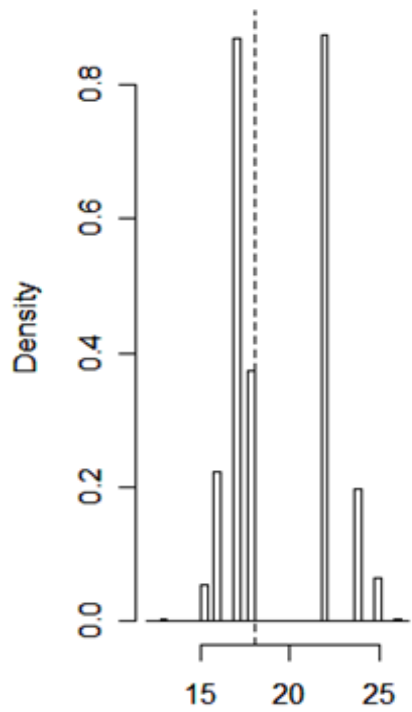

Median *h*-index (Computer science)

**Figure A2: Bootstrap distributions for the median *h*-index**